\documentclass[conference]{IEEEtran}

\usepackage[utf8]{inputenc}

\usepackage{csquotes}

\usepackage[numbers]{natbib}

\usepackage{graphicx}
\graphicspath{{images/}}

\usepackage{amsmath}

\PassOptionsToPackage{hyphens}{url}\usepackage{hyperref}

\usepackage{array}

\usepackage{multirow}

\usepackage{float}

\usepackage{footmisc}

\usepackage{wrapfig}

\usepackage{nameref}

\def\BibTeX{{\rm B\kern-.05em{\sc i\kern-.025em b}\kern-.08em
    T\kern-.1667em\lower.7ex\hbox{E}\kern-.125emX}}

\begin{document}

\title{
    How Do Model Export Formats Impact the Development of ML-Enabled Systems? A Case Study on Model Integration
    {}
}

\author{
    \IEEEauthorblockN{Shreyas Kumar Parida
    }
    \IEEEauthorblockA{
        Vrije Universiteit Amsterdam\\
        Amsterdam, The Netherlands\\
        sparida@student.ethz.ch
    }
    \and
    \IEEEauthorblockN{Ilias Gerostathopoulos}
    \IEEEauthorblockA{
        Vrije Universiteit Amsterdam\\
        Amsterdam, The Netherlands\\
        i.gerostathopoulos@vu.nl
    }
    \and
    \IEEEauthorblockN{Justus Bogner}
    \IEEEauthorblockA{
        Vrije Universiteit Amsterdam\\
        Amsterdam, The Netherlands\\
        j.bogner@vu.nl
    }
}


\maketitle


\begin{abstract}
Machine learning (ML) models are often integrated into ML-enabled systems to provide software functionality that would otherwise be impossible.
This integration requires the selection of an appropriate ML model export format, for which many options are available.
These formats are crucial for ensuring a seamless integration, and choosing a suboptimal one can negatively impact system development, e.g., via increased dependencies and higher maintenance costs.
However, little evidence is available to guide practitioners during the export format selection.

We therefore aim to comprehensively evaluate various model export formats regarding their impact on the development of ML-enabled systems from an integration perspective.
Based on the results of a preliminary questionnaire survey (n=17), we designed an extensive embedded case study with two ML-enabled systems in three versions with different technologies.
We then analyzed the effect of five popular export formats, namely ONNX, Pickle, TensorFlow's SavedModel, PyTorch's TorchScript, and Joblib.
In total, we studied 30 units of analysis (2 systems $\times$ 3 tech stacks $\times$ 5 formats) and collected data via structured field notes.

The holistic qualitative analysis of the results indicated that ONNX offered the most efficient integration and portability across most cases.
SavedModel and TorchScript were very convenient to use in Python-based systems, but otherwise required workarounds (TorchScript more than SavedModel).
SavedModel also allowed the easy incorporation of preprocessing logic into a single file, which made it scalable for complex deep learning use cases.
Pickle and Joblib were the most challenging to integrate, even in Python-based systems.
Regarding technical support, all model export formats had strong technical documentation and strong community support across platforms such as Stack Overflow and Reddit.
Practitioners can use our findings to inform the selection of ML export formats suited to their context.
\end{abstract}


\section{Introduction}
Continuous advancements in computational hardware and the abundant availability of training data have put machine learning (ML)~\cite{Jordan2015} at the center of attention of many industry domains, ranging from power grid management software to autonomous cars or stock trading systems~\cite{dietterich2017steps}.
The impressive predictive and generative abilities of ML models can enable system functionality that was previously impossible.
Therefore, ML models are more and more integrated into ML components~\cite{Lewis2021}, which in turn are then integrated into software systems.
However, these ML-enabled systems also come with new engineering challenges and complexity~\cite{Martinez-Fernandez2022}.
For example, assuring quality attributes like safety and reliability~\cite{gula2020software}, identifying and mitigating new types of technical debt~\cite{Bogner2021}, managing large training data sets~\cite{lwakatare2019taxonomy}, or designing~\cite{Serban2022} and assessing~\cite{Warnett2025} a suitable software architecture for these systems requires considerable expertise and effort.
To tackle these challenges, AI engineering has emerged as a new discipline~\cite{bosch2020ai}, and the synthesis of best practices is slowly but surely making progress~\cite{Amershi2019, Serban2023}.

One such area still in need of guidance is selecting an appropriate \textit{ML model export format}.
This serialization of ML models after their training allows practitioners to store them, typically in the file system~\cite{Toma2024}, and then deploy them outside their training environments.
Since ML-enabled systems are often multi-language systems, with model training typically in Python and inference, e.g., in Java, JavaScript, or Go, interoperability and portability are important characteristics of export formats~\cite{Ahmed2021}.
Most ML frameworks offer a custom export format, but some framework-agnostic options exist as well.
Examples include ONNX, TensorFlow's SavedModel, Pickle, PyTorch's TorchScript, and Joblib.

Recently, a limited number of export formats have been studied regarding their impact on quality attributes like inference duration and energy consumption~\cite{Jacques2024,alizadeh2024green}.
However, the detailed impact of model export formats has not been analyzed from the perspective of \textit{integration}, which \citet{Lewis2021} highlighted as an important and challenging step in the ML engineering process.
Knowing how these formats impact the development and integration of ML-enabled systems is crucial for practitioners when deciding on a format to use.

In this paper, we provide evidence for this impact through an extensive case study with two ML-enabled systems in three technology versions.
The case study design was informed through the results of a preliminary survey about export formats with 17 ML practitioners.
We examined how five model export formats integrate into ML-enabled systems from the perspective of compatibility, system complexity, and integration ease, while collecting data via structured field notes.
The results of our qualitative analysis can guide ML practitioners during the selection of an export format suited to their specific needs.

\section{Background and Related Work}
In this section, we introduce fundamental concepts important for this study, namely the ML engineering process and the chosen export formats, and discuss related work in the area.

\subsection{The Process of Engineering ML-Enabled Systems}
While existing development processes for non-ML software, such as Agile or Waterfall~\cite{rajib2018fundamentals}, are partly applicable for the engineering of ML-enabled systems,
their peculiarities also lead to some notable differences.
While no universally accepted ML engineering process exists, most sources include steps similar to the following ones~\cite{Amershi2019, giray2021software, ranawana2021agile}:

\begin{itemize}
    \item \textbf{Problem Definition}: Define objectives and collect data.
    \item \textbf{Data Preprocessing}: Clean and prepare data.
    \item \textbf{Model Selection, Training, and Evaluation}: Train an appropriate ML model. Evaluate performance to ensure the model meets desired accuracy and other criteria.
    \item \textbf{Software Development}: Develop the ML and non-ML components of the ML-enabled system, e.g., user interfaces, application logic, or prediction components.
    \item \textbf{ML Integration}: As a substage of the software development stage, the integration step incorporates ML models into the ML component of an ML-enabled system. ML models are exported in a suitable format and then imported into the target ML components. Compatibility between the export format and the ML component needs to be ensured, but the format selection can make the integration easier or harder. Additionally, the variety of available formats complicates this choice.
    \item \textbf{System Testing}: Conduct system-wide testing to ensure all components function correctly and meet end-to-end performance requirements.
    \item \textbf{Deployment, Monitoring, and Maintenance}: Deploy the system, monitor its performance, and fix future issues.
\end{itemize}

\subsection{ML Model Export Formats}
Choosing a suitable model export format is vital for effectively integrating ML models into larger systems.
A poor choice is likely to introduce unnecessary dependencies, increase project complexity, and add to the overall integration effort and maintenance~\cite{Sens2024, Jajal2024}.
Additionally, the export format can also impact quality attributes, e.g., in the form of performance bottlenecks, security vulnerabilities, and elevated resource demands.
For instance, improper integration of an ML model can create new attack surfaces, potentially exposing both the model and the system to malicious actors~\cite{lewis2021mismatch, ozkaya2021architecture}.
In conclusion, the importance of selecting the appropriate model export format extends well beyond functionality.
A thoughtful choice in format selection simplifies integration and can directly enhance system performance and security.

\begin{figure}
    \centering
    \includegraphics[width=\linewidth]{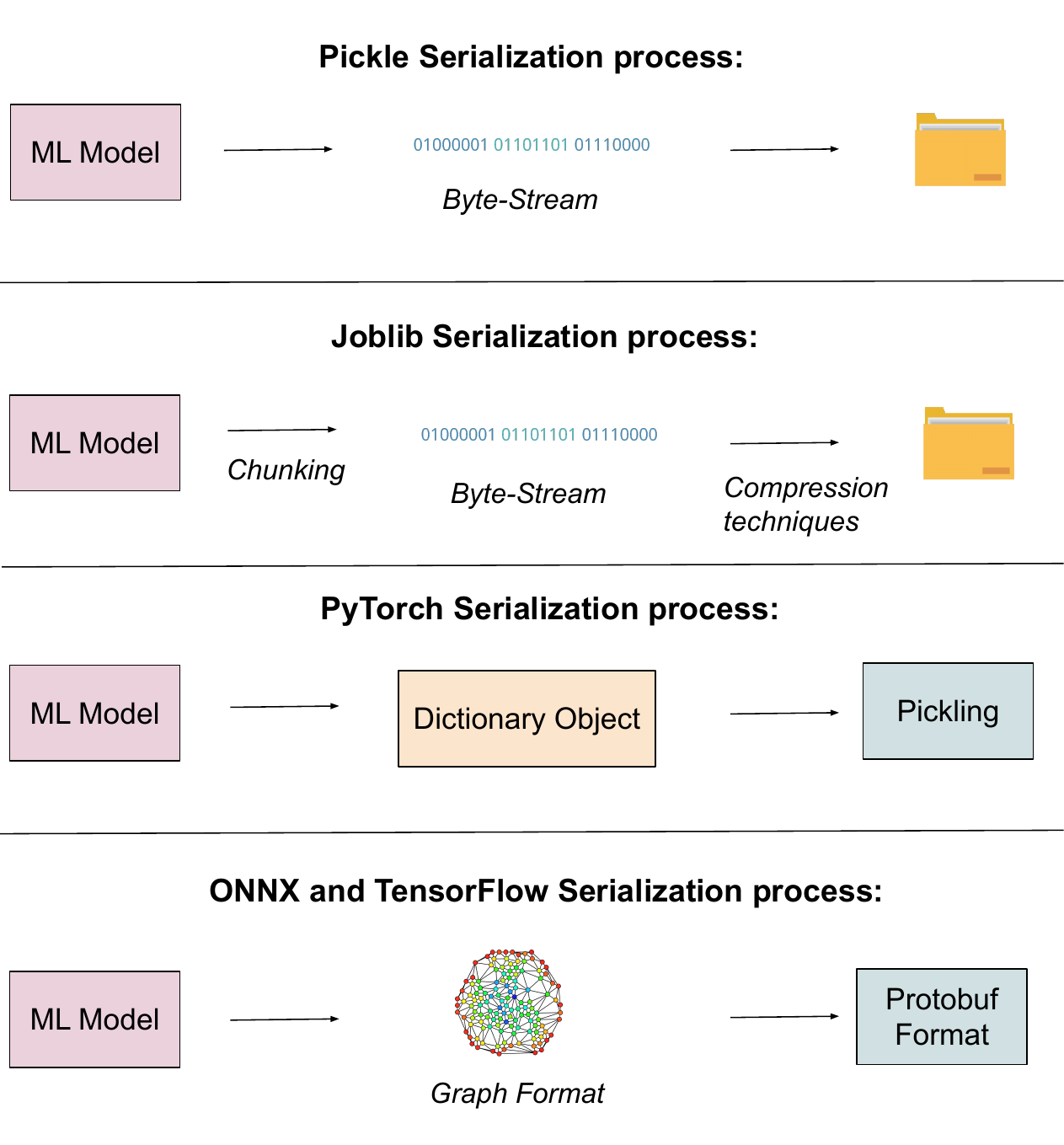}
    \caption{Serialization Process of Model Export Formats}
    \label{fig:export-formats}
\end{figure}

In this study, we evaluated five model export formats regarding their impact on integration, for which understanding their internal mechanisms is essential.
In the following, we introduce these formats. Further justification for why these formats were chosen are provided in later sections.
Additionally, Fig.~\ref{fig:export-formats} visualizes the most important details of the formats.

\subsubsection{ONNX}
The Open Neural Network Exchange (ONNX)\footnote{\url{https://onnx.ai}} format has the goal to make models portable between frameworks and languages.
The export process involves converting the model into a graph, followed by serialization into the Protobuf\footnote{\url{https://protobuf.dev}} format, known for its efficiency.
During import, these steps are reversed~\cite{klingler2023onnx, rai2023onnx}.

\subsubsection{Pickle}
Pickle\footnote{\url{https://docs.python.org/3/library/pickle.html}} is widely used in Python to serialize data structures including ML models into byte streams (often referred to as \enquote{pickling}).
Deserialization (\enquote{unpickling}) converts these streams back into Python data structures.
During unpickling, referenced modules are also required to be imported~\cite{agrawal2024pickling, selvaraj2018pickle}.

\subsubsection{PyTorch's TorchScript}
To export models in PyTorch, TorchScript\footnote{\url{https://pytorch.org/docs/stable/jit.html}} is commonly used, enabling models to be serialized via Pickle in an intermediate format that primarily relies on TorchScript but also involves several PyTorch libraries. TorchScript creates a structured export object that includes the model's tensors, optimizers, and hyperparameters, allowing seamless deserialization and model reconstruction during loading. As this process encompasses multiple components of PyTorch~\cite{pytorch2023saving}, we will refer to the TorchScript export format simply as \enquote{PyTorch}.

\subsubsection{Tensorflow's SavedModel}
TensorFlow offers a custom export format called SavedModel\footnote{\url{https://www.tensorflow.org/guide/saved_model}}.
This format is similar to ONNX, where the model is converted to a graph and serialized into Protobuf.
While this is supposed to allow model transfer across TensorFlow implementations in different programming languages~\cite{tensorflow2023savedmodel}, the difference to ONNX is that no cross-framework portability is supported.

\subsubsection{Joblib}
Joblib\footnote{\url{https://joblib.readthedocs.io/en/stable}} serializes Python objects similarly to Pickle but breaks ML models into smaller compressed chunks for parallel serialization, which is supposed to offer speed and memory benefits~\cite{joblib2023github}.

Moreover, several model export formats come with a \textit{runtime module}.
These modules play a crucial role in integrating models into ML-enabled systems by handling tasks like model loading, dependency management, computational resource allocation, and inference execution.
They also facilitate data handling and ensure smooth interaction between the model and the system.
Official runtime modules, such as the ONNX runtime or PyTorch's JIT compiler, are developed by the creators of the model formats and offer optimized performance and reliability.
Unofficial runtime modules, developed by third-party contributors, can provide additional flexibility, but may lack the stability and support of official versions.
Generally, official runtime modules are preferred due to their seamless integration and direct support from the format developers~\cite{saturn2023exporting}.

\subsection{Related Work}
Despite the undisputed importance of the integration stage of ML-enabled systems, very few studies exist on the topic of model export formats, highlighting a gap this research aims to fill.
\citet{shridhar2020onnx} analyzed ONNX's stability for integrating deep learning models with Julia-based systems, while \citet{olston2017tensorflow} shared Google's largely positive experiences with TensorFlow, acknowledging potential bias due to their involvement in TensorFlow's development.
\citet{alizadeh2024green} showed ONNX's performance advantages in runtime modules, while \citet{peng2020first} emphasized the impact of integration quality on performance in autonomous driving systems, advocating for more extensive testing.
These findings collectively indicate that optimal model export format selection can significantly enhance development and system performance, whereas suboptimal choices may reduce efficiency and effectiveness.

More broadly, the topic of ML model integration in ML-enabled systems has been recently studied via mining open-source repositories.
\citet{nahar2024product} found that (i) about half of the products integrate third-party, i.e., pre-trained instead of self-trained, ML models, and (ii) the majority of products integrate several mostly independent models.
\citet{Sens2024} categorized the integration of ML models in four distinct patterns: \textit{alternative} (several models for the same task), \textit{independent} (different models for different tasks), \textit{sequential} (cascading model invocations), and \textit{joining} (results of different models are combined by another model or code). 
They also noted that loading ML models can become a complex configuration and assembly problem. 
\citet{Toma2024} found that ML models are loaded from files for three reasons: for evaluation or inference (most common case), for resuming training, and for transforming them to different formats.
The latter happens, e.g., for deploying ML models in a target environment or for reducing their size.
However, none of these three studies include a discussion of the concrete export formats used in the analyzed repositories.
Our study complements this body of knowledge with best practices for model export format selection, which is part of every model integration process.

Recent works have also focused on the timely problem of reusing and re-engineering ML models, in particular for deep learning (DL).
\citet{Jiang2024} identified the portability of DL operators  and the complex data pipelines as two main challenges in re-engineering DL models.
\citet{Davis2023} categorized DL model reuse into conceptual reuse, adaptation reuse, and deployment reuse.
DL interoperability is a primary challenge in the latter, and can be tackled by using standardized representations such as ONNX or model converters~\cite{Jajal2024}.
The choice of model export format, which can be informed by our study, clearly also affects the interoperability of the resulting ML/DL model.
In summary, the concrete integration impact of different export formats has so far been neglected by existing studies.

\section{Study Design}
To fill the identified research gap, we first conducted a preliminary questionnaire survey~\cite{Kitchenham2008} with ML practitioners to find out about popular formats, what they are used for, and which factors influence their selection.
The results of this survey then informed the design of a case study~\cite{runeson2008guidelines}.
For transparency and reproducibility, we make our study artifacts available online.\footnote{\url{https://doi.org/10.6084/m9.figshare.27613212}}
Overall, our research into the topic of ML model export formats was guided by the following research questions:

\textbf{RQ1}: How effectively and efficiently can existing ML model export formats be integrated into ML-enabled systems?

With this RQ, we wanted to understand how compatible each export format is with different technology stacks, how much effort is required to accomplish the integration, and which challenges can arise during it.

\textbf{RQ2}: To what extent is the integration of ML model export formats impacted by the scale and complexity of the underlying ML model?

Since ML use cases can be of different complexity, we wanted to study how this influences the integration of different formats.
It could be possible that some formats are easy to integrate with small, simple models, but not with large, complex deep learning models.

\textbf{RQ3}: What level of technical support exists for each model export format?

Using a certain export format can be made much easier via high-quality documentation, as well as community activity on platforms like Stack Overflow in case of problems not solvable via documentation.
We wanted to understand if this support is different for the various export formats.

\subsection{Preliminary Questionnaire Survey}
To ground our case study in industry-relevant model export formats, we conducted a short online questionnaire survey that was distributed to various ML practitioners via personal email contacts, Kaggle forums, LinkedIn groups and ML subreddits. 
Apart from providing data to make an appropriate selection of model export formats, the questionnaire also aimed to acquire useful information to answer the research questions and inform the case study design.
In addition to some demographic information, the survey contained the following questions:

\begin{itemize}
    \item Which ML model export format(s) do you use regularly? \textit{(multiple choice question with \enquote{Other: ...} option)}
    \item For what primary purpose do you use the ML model export format(s)? \textit{(multiple choice question with \enquote{Other: ...} option)}
    \item What factors influence your choice of ML model export format(s)? \textit{(multiple choice question with \enquote{Other: ...} option)}
    \item Have you encountered any challenges with your current model export format(s)? \textit{(free-text question)}
\end{itemize}

Overall, 17 participants responded to our survey.
They can be split into two subgroups: 12 ML practitioners (half with 5+ years of experience, half with 1-3 years) and 5 students.
Participants spanned industry domains like Software \& IT, Finance, Education, and Health.
The results indicated that ONNX was the most commonly used model export format among participants, with five users. PyTorch’s TorchScript also saw wide usage, with four users favoring it. Other formats, such as Pickle (3 users), Joblib (2 users), and TensorFlow's SavedModel (2 users), saw moderate use, while JSON was the least utilized, with only one user.
Preferences were similar across both subgroups, likely reflecting ONNX's broad compatibility across frameworks and PyTorch's strong presence in deep learning applications.
Regarding the main factors influencing the choice of export format (see Table~\ref{table:model_export_factors}), the most common reported reasons were ease of use (12 mentions) and compatibility (8).
One participant notably mentioned \enquote{Security Features}, but provided no further details.

\begin{table}[ht!]
    \centering
    \caption{Factors Influencing the Choice of Machine Learning Model Export Formats}
    \begin{tabular}{lr}
    \textbf{Influencing Factor} & \textbf{\# of Mentions} \\
    \hline
    \hline
    Ease of Use & 12 \\
    Community Support and Documentation & 7 \\
    Compatibility with Deployment Environments & 4 \\
    Performance Optimization & 4 \\
    Scalability & 2 \\
    Security Features & 1 \\
    \hline
    \hline
    \end{tabular}
    \label{table:model_export_factors}
\end{table}

In response to the free-text question on challenges with model export formats, only one detailed reply was received, likely due to the added effort required for open-ended answers.
A practitioner with over five years of experience suggested suboptimal documentation and difficult inference optimization.

\begin{figure*}[ht]
    \centering
    \includegraphics[width=1\linewidth]{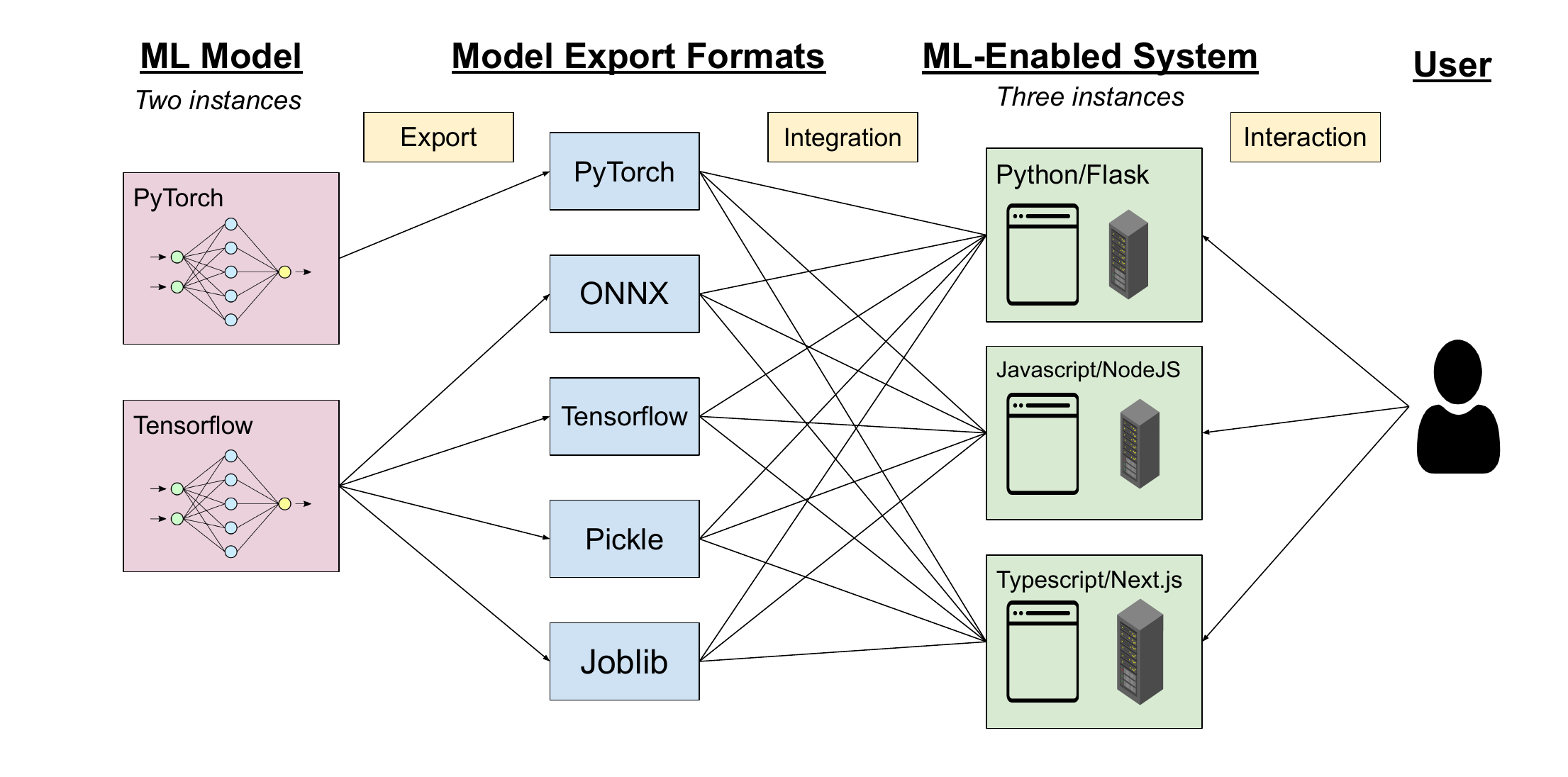}
    \caption{Development Process for Each of the Two ML-Enabled Systems (Number Predictor and Sentiment Analysis Tool)}
    \label{fig:development-process}
\end{figure*}

\subsection{Case Study Design}
Using the survey results as well as the formats reported in related work and gray literature~\cite{saturn2023exporting}, we chose the following five export formats for our case study:
ONNX, Pickle, SavedModel from TensorFlow (will be referred to as TensorFlow), TorchScript (will be referred to as PyTorch due to its respective export / import covering multiple libraries of PyTorch beyond TorchScript), and Joblib.
To develop and export the models in these formats, we mostly used the TensorFlow framework\footnote{\url{https://www.tensorflow.org}}, which is one of the most popular choices in this space~\cite{kaggle2022survey} and provides sufficient functionalities for developing neural networks.
The only exception was the TorchScript format, which is more suited to development using the PyTorch framework\footnote{\url{https://pytorch.org}}.

This selection covered popular frameworks and formats with different philosophies, while still being manageable for the implementation part.
To answer our research questions, we conducted an \textit{embedded} case study~\cite{runeson2008guidelines}, i.e., we had multiple units of analysis per case.
For diversity but also to have different model complexities for RQ2, we decided to develop and study two ML-enabled systems (two \textit{cases}), each with its own ML model to be integrated.

The first ML-enabled system was a simple number predictor, chosen for its straightforward design and ease of implementation.
Given three numbers, it predicted the next using a three-layer dense network (64, 32, 1 neurons) with ReLU and linear activations.
The second system was a sentiment analysis tool for movie reviews, capable of classifying a review as positive or negative.
The respective model was a neural network trained for sentiment analysis.
The sentiment analysis use case was chosen due to the abundant literature on the topic~\cite{lu2023sentiment, hu2004mining, huq2017twitter} and its increased complexity compared to the number predictor.
Text-based sentiment classification involves additional preprocessing for user inputs, as training the model on raw text input is typically inefficient~\cite{lu2023sentiment}.
Instead, the model was trained on embedding vectors, which converted the text into numerical patterns that the model could more easily process.
However, this required that all user inputs undergo the same preprocessing steps, i.e., embedding vector conversion, adding to the complexity of the system.
Additionally, this model was considerably larger and deeper than the number predictor model.
It had three dense layers (256, 128, and 1 neurons), totaling over 300,000 parameters to process complex text patterns.
We wanted to study how the different export formats would react to this increased complexity.

Within each of the two cases (number predictor and sentiment analysis tool), we introduced multiple units of analysis based on different technology stacks.
By using different programming languages and application frameworks, we were able to test interoperability and portability of the different export formats. 
Since both systems work well as web-based applications, we selected different web technologies to implement them.
Various surveys~\cite{stackoverflow2023survey, statista2023webframeworks} indicated that JavaScript, Typescript, and Python are the most popular languages for web application development.
Therefore, we decided to use these three languages and to select popular web frameworks for each.
We chose Node.js's \texttt{http} module\footnote{\url{https://nodejs.org/api/http.html\#http}} for JavaScript, Next.js\footnote{\url{https://nextjs.org}} for TypeScript, and Flask\footnote{\url{https://flask.palletsprojects.com/en/3.0.x}} for Python.

Overall, our embedded case study design involved a total of 30 units of analysis, i.e., 2 systems $\times$ 3 tech stacks $\times$ 5 formats.
This led to decent diversity, allowing a more balanced analysis of the integration impact of the five export formats.

\subsection{Study Execution \& Data Collection}
The development of the number predictor involved two main stages.
The first stage was training the ML model, once using PyTorch and once using TensorFlow.
Once trained, the model was exported in the various formats.
The second stage focused on building the full ML-enabled system, which included creating the necessary components like the user interface, domain logic, and other backend features.
The system was built in the three mentioned versions, each following a different programming language and web application framework.
Each of these versions was then integrated with the ML model, resulting in a fully functional, end-to-end ML-enabled number predictor.

The process for the sentiment analysis tool was very similar, with the first stage involving again the training of the ML model with subsequent export.
Afterward, the complete ML-enabled system was developed in the three versions, which were then integrated with the ML model. This resulted in a fully functional, end-to-end sentiment analysis tool for movie reviews.
Fig.~\ref{fig:development-process} illustrates the above-mentioned development process.

While following the above process, we used structured field notes~\cite{seaman1999qualitative} to document the followed steps and encountered challenges.
These notes focused on capturing the experience of integrating the ML model into the ML-enabled systems, explicitly recording both issues and their solutions or workarounds when available.
Our developed field note template had the following structure:

\begin{itemize}
    \item A title identifying the system, model export format, and web application framework.
    \item Description of the integration process, main steps, encountered challenges, and their solutions.
    \item A subjective ordinal integration score with these levels:
    \begin{itemize}
        \item Seamless integration: ++
        \item Minor issues during integration: +
        \item Major issues during integration: -
        \item Integration is not possible or highly challenging: \texttt{-{}-}
    \end{itemize}
    \item \textit{(Only for some integrations to prevent duplicate data)} During the investigation of encountered problems, but also at least once per format: comments on how much community support was available for each model export format on major tech forums such as Stack Overflow, smaller community pages, and official documentation.
\end{itemize}

\begin{figure}[H]
    \centering
    \includegraphics[width=1\linewidth]{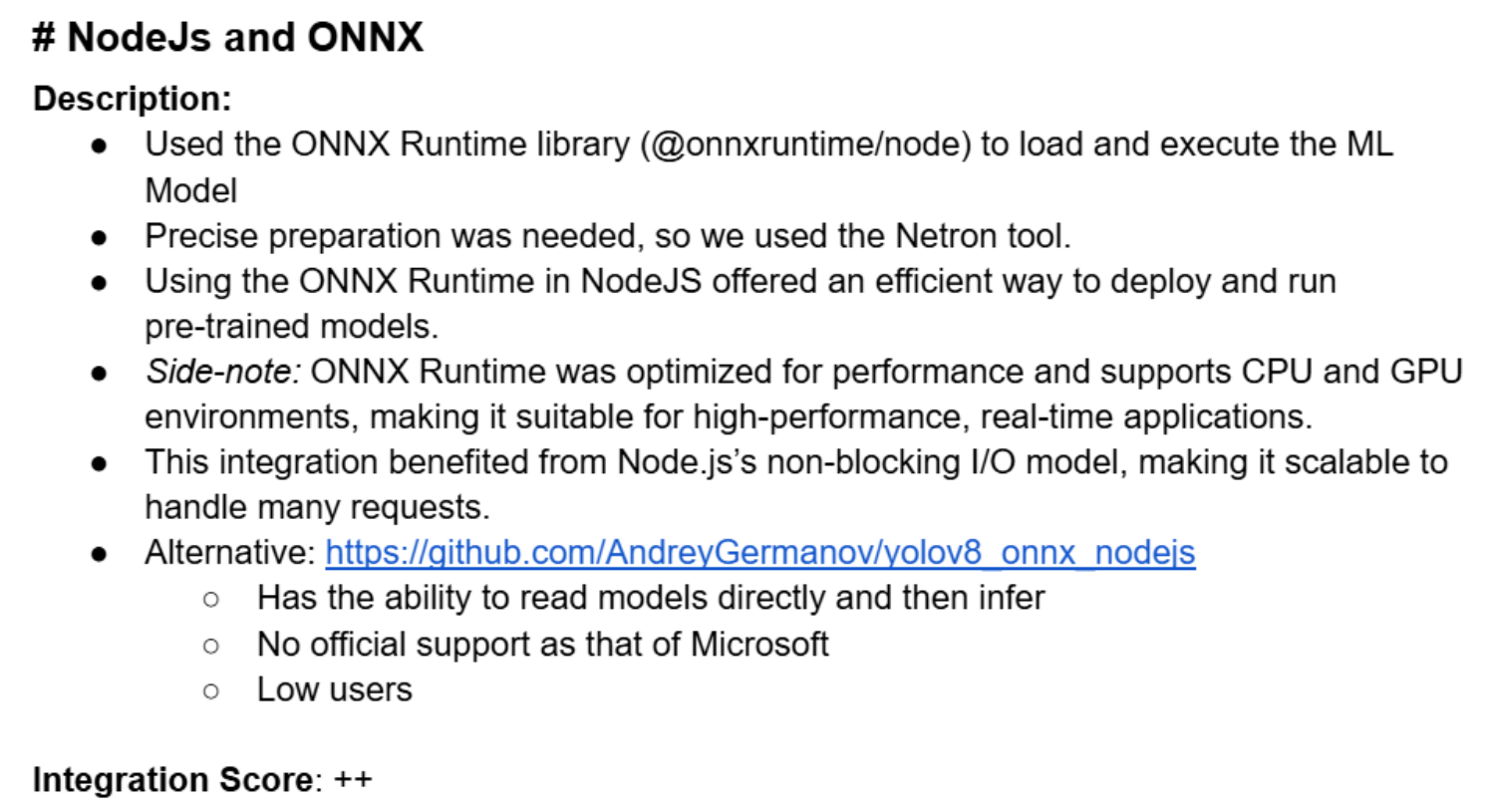}
    \caption{Example of Structured Field Notes}
    \label{fig:field-note-example}
\end{figure}

This structured approach ensured a thorough and standardized assessment of how the various model export formats integrated with the different ML-enabled systems.
In total, we created 30 field notes during the study.
Fig.~\ref{fig:field-note-example} illustrates a concrete field-note example.

During the case study, we also collected data on the community sizes of each Model Export Format, focusing on GitHub, Stack Overflow, and Reddit.
The primary data collection method involved visually inspecting each respective community's membership counts.

\subsection{Data Analysis}
After finishing the study execution, we holistically analyzed all field notes to identify the main information for answering the research questions. 
Using lightweight thematic analysis~\cite{seaman1999qualitative}, we noted down major themes such as encountered challenges, and also aggregated the experiences for each export format into a more cohesive judgment. 
To this end, the six field notes per format were analyzed together to synthesize their overall impression, with separate takeaways per research question. To enhance validity, we cross-checked the synthesized results within the research team, allowing for feedback to ensure consistency and reduce individual bias. 
Finally, we compared the results with our survey data and related work, and wrote a results text per export format. A summarizing rating table outlining the integration ease of each format per web app framework was created and discussed in the research team.

\section{Results}
In this section, we present the case study results according to the research questions.
Tables~\ref{table:model_export_sizes} and~\ref{table:model_export_sizes_sentiment} summarize the exported models and their sizes per format.

\begin{table}[ht!]
    \centering
    \caption{Size of Different Model Export Formats for Number Predictor}
    \begin{tabular}{llr}
    \textbf{Model Export Format} & \textbf{File Extensions} & \textbf{Size} \\
    \hline
    \hline
    ONNX & \texttt{.onnx} & 10.9 kB \\
    TensorFlow\footref{fn:tensorflow} & 3x \texttt{.pb}, 1x \texttt{.index}, 1x \texttt{.data} & 142.8 kB \\
    PyTorch & \texttt{.pt} & 29.8 kB \\
    Pickle & \texttt{.pkl} & 63.7 kB \\
    Joblib & \texttt{.joblib} & 66.3 kB \\
    \hline
    \hline
    \end{tabular}
    \label{table:model_export_sizes}
\end{table}

\footnotetext{A TensorFlow model is exported and imported into a system as a ZIP archive, i.e., developers typically handle it as a single file. The details shown in this table are based on the contents after extracting the ZIP archive.\label{fn:tensorflow}}

\begin{table}[ht!]
    \centering
    \caption{Size of Different Model Export Formats for Sentiment Analysis Tool}
    \begin{tabular}{llr}
    \textbf{Model Export Format} & \textbf{File Extensions} & \textbf{Size} \\
    \hline
    \hline
    ONNX & \texttt{.onnx} & 37.3 MB \\
    TensorFlow\footref{fn:tensorflow} & 3x \texttt{.pb}, 1x \texttt{.index}, 1x \texttt{.data} & 31.6 MB \\
    PyTorch & \texttt{.pt} & 10.4 MB \\
    Pickle & \texttt{.pkl} & 62.2 MB \\
    Joblib & \texttt{.joblib} & 62.2 MB \\
    \hline
    \hline
    \end{tabular}
    \label{table:model_export_sizes_sentiment}
\end{table}

\subsection{Ease of Integration (RQ1)}
For RQ1, we wanted to explore the required effort and challenges of integrating each model export format into the ML-enabled systems.
We will discuss most export formats individually and then present a holistic summary at the end.
Regarding terminology, we refer to any ML model exported into a specific export format as an \enquote{X model}, where X is the name of the respective format.
For example, a model exported in the ONNX format will be called an \enquote{ONNX model}.

\subsubsection{ONNX and TensorFlow}\hfill\\
We present the results of integrating the ONNX and TensorFlow models together due to their similarities.

\textbf{Python}: Integrating the ONNX and TensorFlow models with the ML-enabled system developed in Python was straightforward.
The Python environment allowed for the convenient import of the official ONNX\footnote{\url{https://onnxruntime.ai}} and TensorFlow\footnote{\url{https://github.com/tensorflow/runtime}} runtime modules.
Thus, once the models were loaded using the runtime modules, we could call the models on demand, enabling seamless integration for prediction or classification tasks.
We also briefly searched for unofficial runtime modules worth investigating but did not find any popular candidates, which speaks for the quality of the official ones.

However, even with such a smooth integration, we encountered a problem with setting up the ONNX runtime module.
For this to work correctly, various details of the underlying neural network were required, such as the model architecture, input-output dimensions, and specific configurations.
This led to a problematic circular dependency:
the ONNX runtime module was required to operate a black-box ONNX model in Python, but details of the underlying neural network had to be provided for this to work.
However, from the perspective of a model integrator without access to the model training details, these attributes could only be acquired via the ONNX runtime module.
While we tried several approaches to address this challenge, the simplest approach was to use an external visualization tool such as Netron\footnote{\url{https://github.com/lutzroeder/netron}}.
The Netron tool parsed the ONNX model and displayed its underlying neural network in a graphical format.
As such, the relevant information could be easily extracted and was used to initialize the ONNX runtime module, which completed the integration.
The above-mentioned problem did not occur during the integration of the TensorFlow model.

\textbf{JavaScript / TypeScript}: The integration with JavaScript and TypeScript is discussed together, as both these frameworks had a very similar integration process.

The ONNX model integration with the JavaScript and TypeScript systems was similar to that of Python.
The official runtime module in the respective programming language was first imported and then initialized using the data from the Netron tool, yielding an error-free integration process.

For TensorFlow, we first needed to integrate a working runtime module for JavaScript and TypeScript.
As such, we decided to import TensorFlow.js\footnote{\url{https://github.com/tensorflow/tfjs}} into both systems, which is an official runtime module.
However, a prerequisite for using this runtime module was that the TensorFlow model had to be re-trained and re-exported using the Python module TFJS\footnote{\url{https://pypi.org/project/tensorflowjs}}, a custom variation of the TensorFlow ML framework.
The resulting trained model was exported and could then be easily integrated with the TensorFlow.js runtime module.
So, despite TensorFlow's supposed cross-language capabilities, it was still necessary to plan accordingly for such portability.

\textbf{Analysis}: Integrating the ONNX and TensorFlow models into the ML-enabled systems went fairly efficiently and without substantial challenges.
The efficient integration was mainly made possible by the official runtime modules. 
However, the ONNX runtime module required a visualization tool to help initialize the runtime module quickly.
While the TensorFlow runtime module did not require the same external initialization details, these embedded neural network details came at a cost: the TensorFlow model was far larger than the ONNX model (approximately 15 times, see Table~\ref{table:model_export_sizes}).
Lastly, we needed to retrain and export the TensorFlow model with a different library to allow its import into the TensorFlow.js runtime module.

\subsubsection{PyTorch}\hfill\\
\textbf{Python}: Integrating the PyTorch model into the Python-based system required importing the \texttt{torch} module\footnote{\url{https://pytorch.org/docs/stable/library.html}}.
This module contained all PyTorch-related tools.
However, the main tool of interest for this integration was PyTorch's JIT compiler\footnote{\url{https://pytorch.org/docs/stable/jit.html\#module-torch.jit}}, which allowed interpreting, executing, and parsing the PyTorch model.
Unlike the ONNX runtime, the JIT compiler did not require any details of the underlying neural network for setup.
The integration process was therefore very efficient.  

\textbf{JavaScript / TypeScript}:  Unfortunately, no official runtime modules that provide similar functionalities to the JIT compiler exist for JavaScript or TypeScript.
However, various non-native runtime modules promise smooth integration between JavaScript / TypeScript and the PyTorch model.
Due to factors such as insufficient compatibility, potential security risks, or minimal documentation, we did not find any suitable candidate, though.
For example, Transformers.js\footnote{\url{https://www.npmjs.com/package/@xenova/transformers}} can convert PyTorch models into ONNX models before running them, but it only supports Transformer-based models.

As a consequence, we had to establish a Python subprocess within Node.js as a workaround\footnote{\url{https://nodejs.org/api/child_process.html\#child-process}}.
This enabled us to utilize standard Python capabilities for model inference.
We could therefore use similar code as for the PyTorch model integration with Python, which completed the integration successfully, albeit in a convoluted way.

\textbf{Analysis}: PyTorch's JIT compiler aided greatly in the integration process, automating many tasks such as model
loading, dependency management, and inference execution.
This allowed us to save effort during the integration with the Python-based system.
On the other hand, the absence of a compatible JavaScript runtime module substantially decreased the integration quality for the non-Python systems.
Spawning a Python subprocess from Node.js was a successful workaround to complete the integration, but it definitely had at least some negative impact on quality attributes like maintainability and performance efficiency.

\subsubsection{Pickle and Joblib}\hfill\\
Due to their similarities in mechanisms and outcomes, we will discuss Pickle and Joblib together.

\textbf{Python}: Despite their Python origin, integrating the Pickle and Joblib models with the ML-enabled system developed in Python presented many challenges.
Pickle is a built-in Python function, while we imported the official runtime module for Joblib\footnote{\url{https://github.com/joblib/loky}}.
However, numerous errors occurred during model deserialization, most of them due to the runtime module's inability to parse the model weights.
We tried the following (unsuccessful) solutions to fix this:

\begin{itemize}
    \item Explicitly adding the model weights to the respective Pickle and Joblib exports.
    \item Separating the model structure and weights into two different Pickle / Joblib files.
    \item Training the ML model using only weight attributes supported by Pickle or Joblib (according to their official documentation).
\end{itemize}

When none of this worked, we instead chose a very different approach, namely switching the model training and export from TensorFlow to a different ML framework.
As it is much more related to the Pickle and Joblib formats, we opted for scikit-learn\footnote{\url{https://scikit-learn.org}} (\texttt{sklearn}), one of the oldest ML frameworks, which is also decently popular~\cite{kaggle2022survey}.
This switch indeed solved the problem, i.e., after training and exporting with \texttt{sklearn}, both the Pickle and Joblib model could be integrated into the Python application in a straightforward and error-free way.

\textbf{JavaScript / TypeScript}: Integrating Pickle and Joblib models with the JavaScript and TypeScript systems faced the challenge that no official JavaScript-based runtime modules exist.
Since initial explorations into unofficial runtime modules were unsuccessful, we started investigating if JavaScript runtime modules existed for the \texttt{sklearn} framework.
While some candidates are available, such as scikit.js\footnote{\url{https://www.npmjs.com/package/scikitjs}}, we found all of them to be unreliable due to low usage, lack of documentation, and missing functionalities.
We therefore followed the same Python subprocess approach as for the PyTorch models.
This resulted in a similarly effective yet inelegant integration, which also relied on switching from TensorFlow to \texttt{sklearn}.

\textbf{Analysis}: Integrating Pickle and Joblib models presented varying challenges for all three systems, even the Python-based one.
The successful integrations were largely achieved through alternative strategies and tools.
As the serialization from TensorFlow did not include the model weights, switching to \texttt{sklearn} was a prerequisite for successfully using Pickle and Joblib.
Additionally, the same problematic subprocess workaround had to be used for the JavaScript-based systems.

\subsubsection{Summary}\hfill\\
Table~\ref{tab:rating-summary} aggregates the general ease of the described integration processes into one rating per model export format and system language.
The ratings go from \texttt{-{}-}, indicating a highly challenging integration, to \texttt{++}, indicating a very simple and smooth one.
Furthermore, these integration scores are mostly provided in relation to the respective technology stacks.
The results should therefore ideally be interpreted column-wise, not row-wise.

\begin{table}[H]
    \centering
    \caption{Aggregated Ease of Integration Score per Model Export Format and System Language (from -{}- to ++)}
    \begin{tabular}{p{2cm}p{1cm}p{1cm}p{1cm}}
        \multirow{2}{3cm}{\textbf{ML Model \\ Export Formats}} & \multicolumn{3}{c}{\textbf{System Language}} \\
        & \textbf{Python} & \textbf{TypeScript} & \textbf{JavaScript} \\
        \hline
        \hline
        ONNX & + & ++ & ++ \\
        Pickle & - & -{}- & -{}- \\
        PyTorch & ++ & - & - \\
        TensorFlow & ++ & + & + \\
        Joblib & - & -{}- & -{}- \\
        \hline
        \hline
    \end{tabular}
    \label{tab:rating-summary}
\end{table}

\subsection{Impact of Model Complexity (RQ2)}
For RQ2, we wanted to understand if model complexity impacted the integration process of these export formats.
The answers were derived by comparing how each format fared for the two systems, specifically if the integration was more difficult for the sentiment analysis tool.
The model of this system was more complex in two independent areas: a) regarding its size and depth, and b) regarding the amount of required preprocessing.
Apart from this, we tried to keep the integration steps between the model export formats and ML-enabled systems as close as possible to those for the simpler number predictor.
We made observations in several areas.

\subsubsection{Impact of Model Size and Depth}\hfill\\
Across all model export formats, no problems or substantial inefficiencies arose as the exported ML model grew larger and deeper.
As such, the more complex model did not substantially affect the integration process.
All used export formats are pretty mature and therefore presumably fairly robust towards model complexity.
However, we also did not experiment with a large variety of model sizes, especially not extremely large generative models.
It is possible that some export formats would have reacted differently under such conditions.

\subsubsection{Preprocessing Impact}\hfill\\
The ONNX, Pickle, and Joblib export formats each followed a similar procedure for handling the increased preprocessing of model inputs.
During model training and export, the preprocessing logic was exported as a second file, with the first file containing the ML model.
To complete the integration, the runtime module in each system was updated to accommodate this preprocessing logic, ensuring that it was correctly applied to all inputs before inference.

In contrast to the above, the TensorFlow model export format offered the possibility to integrate both the preprocessing logic and the ML model in the same file, making the integration considerably simpler.
Consequently, no further modifications were required to the ML-enabled systems to support the additional preprocessing.
However, the downside of this convenience was a tight coupling between ML model and preprocessing logic, which now prevented adjusting the preprocessing without replacing the model.
Since preprocessing and ML model usually co-evolve, this may be negligible in many cases, but it is still something to consider.

For the PyTorch integration, we initially trained the ML model using the PyTorch framework.
However, including the text preprocessing logic directly in PyTorch caused run-time errors, such as \texttt{unsupported layer type for conversion}.
To work around this, we trained an equivalent model using \texttt{sklearn} to handle preprocessing separately.
We then manually recreated the model architecture in PyTorch using the weights and structure from the \texttt{sklearn} model.
This allowed us to export both the model and preprocessing logic in a single file, similar to TensorFlow’s SavedModel.
Nevertheless, the required workaround highlighted the limitations for models with complex preprocessing needs.
Finally, as shown in Table~\ref{table:model_export_sizes_sentiment}, for the complex model, formats like ONNX, Pickle, and Joblib produced larger files than TensorFlow's SavedModel, suggesting that these formats may incorporate extra parameters as model complexity grows, while TensorFlow may optimize for size.

\subsection{Available Technical Support (RQ3)}
All model export formats provide comparable technical support, including detailed documentation and troubleshooting guidance.
Users can explore GitHub repositories for updates and issue resolutions, while platforms like Stack Overflow and Reddit offer additional community support across all formats.

\begin{figure}[H]
    \centering
    \includegraphics[width=\linewidth]{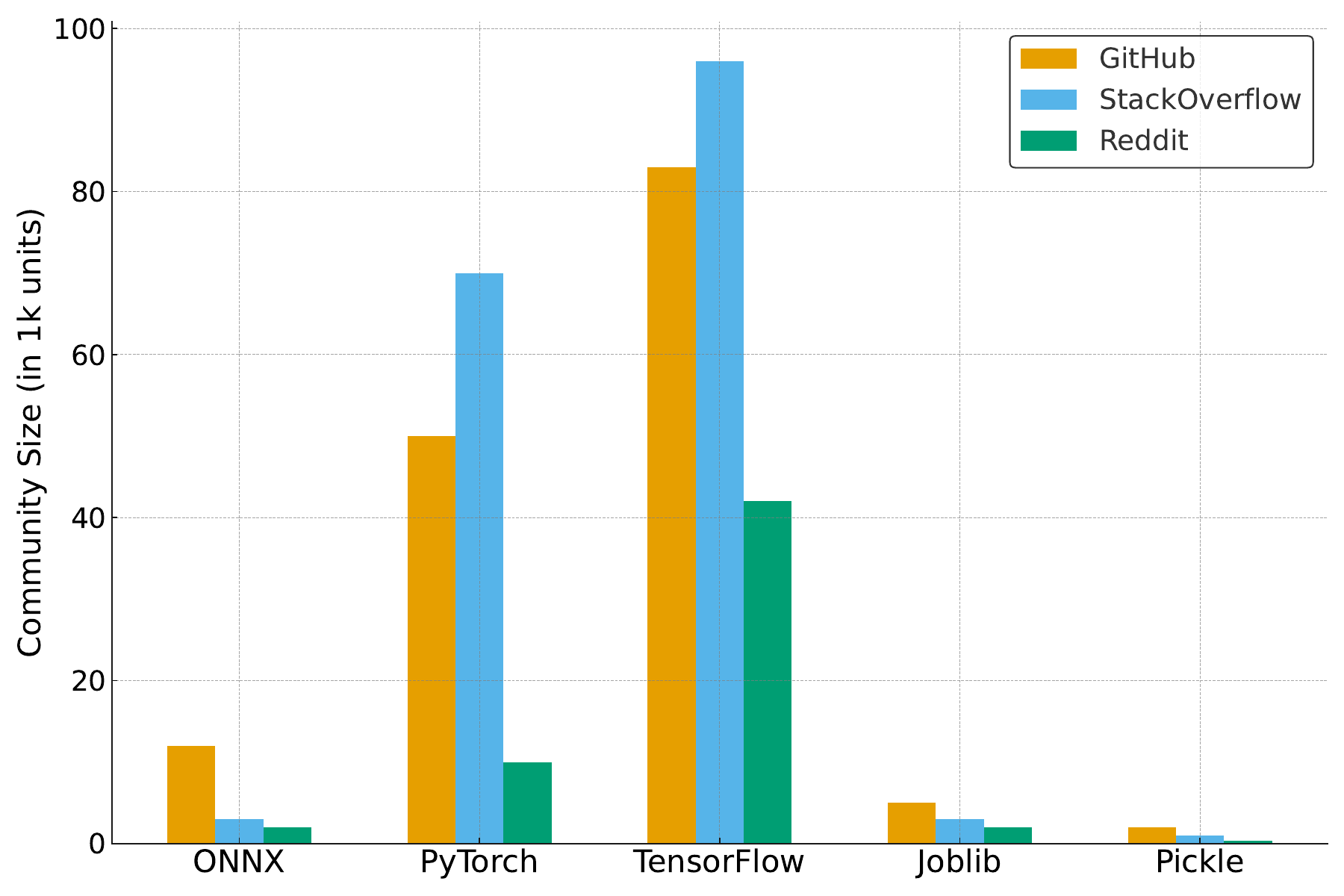}
    \caption{Community Size of Different Model Export Formats}
    \label{fig:community-size}
\end{figure}

While TensorFlow and PyTorch have the largest communities (see Fig.~\ref{fig:community-size}), they also cover many topics beyond export formats.
In contrast, the ONNX, Joblib, and Pickle communities focus more specifically on export-related issues, though community size alone may not fully indicate the quality of the support.
Despite one of our survey participants reporting suboptimal documentation as a challenge, we generally experienced no noticeable differences regarding the available support quality across export formats.
This seems to indicate that all of them are supported by a mature community ecosystem.
We documented identified support resources like documentation, GitHub repositories, or tutorials during our case study.
A list of these resources is available in our replication package.

\begin{table*}[ht!]
    \centering
    \caption{Usage Guidance per Model Export Format (if switching the format is not really an option)}
    \label{tab:usage-guidance}
    \begin{tabular}{lp{15.5cm}}
        \textbf{Export Format} & \textbf{Recommendations \& Tips} \\
        \hline
        \hline
        \vspace{-0.3cm}
        TensorFlow &
        \vspace{-0.25cm}
        \begin{enumerate}
            \item For Python-based systems, use the official TensorFlow runtime module.
            \item For JavaScript / TypeScript applications, use the TensorFlow.js runtime. Re-train and export the model using TFJS (Python).
            \item Preprocessing logic can be embedded into the ML model, leading to a single export file.
        \end{enumerate} \\
        \vspace{-0.3cm}
        ONNX & 
        \vspace{-0.25cm}
        \begin{enumerate}
            \item Use the official ONNX runtime module in the language of your system.
            \item Use a visualization tool like Netron to retrieve the necessary information to initialize the runtime.
            \item Preprocessing logic requires a separate file to be exported alongside the ML model.
        \end{enumerate} \\
        \vspace{-0.3cm}
        PyTorch & 
        \vspace{-0.25cm}
        \begin{enumerate}
            \item For Python-based systems, use the official PyTorch runtime module.
            \item For JavaScript / TypeScript applications, avoid using TorchScript and instead export to ONNX. If TorchScript is unavoidable, spawn a Python subprocess from Node.js for inference as a workaround.
            \item Not recommended for models with complex preprocessing logic.
        \end{enumerate} \\
        \vspace{-0.3cm}
        Pickle / Joblib & 
        \vspace{-0.25cm}
        \begin{enumerate}
            \item Train the ML model using the \texttt{sklearn} framework for best compatibility.
            \item For Python-based systems, use \texttt{sklearn} and the official runtime module for inference.
            \item For JavaScript / TypeScript applications, avoid using Pickle / Joblib and instead export to ONNX. If Pickle / Joblib are unavoidable, spawn a Python subprocess from Node.js for inference as a workaround.
            \item Preprocessing logic requires a separate file to be exported alongside the ML model.
        \end{enumerate} \\
        \hline
        \hline
    \end{tabular}
\end{table*}

\section{Implications}
Our integration experiences across different technology stacks highlighted key insights into compatibility challenges and associated efforts, which can support practitioners in selecting and using these formats.

\textbf{Python Integration}: Our integration ratings showed the highest scores for the Python-based system with the TensorFlow and PyTorch export formats.
These scores were mostly attributed to their official Python runtime modules that required minimal configuration.
Both of these formats are therefore good choices for practitioners if the inference happens in a Python component.
The ONNX integration was moderately smooth, but required model details accessible through tools like Netron.
However, once identified, this step did not significantly increase the integration effort.
Conversely, Pickle and Joblib export formats encountered serialization challenges, even in a Python environment.
Both formats therefore cannot be recommended if TensorFlow or PyTorch are the used ML frameworks.
A potential valid reason for using them might be an existing legacy ML training project relying on \texttt{sklearn}.
Because both export formats are historically tied to this framework, they showed good compatibility there.
    
\textbf{JavaScript and TypeScript Integration}: The JavaScript and TypeScript systems exhibited similar integration outcomes.
The Pickle, Joblib, and PyTorch export formats could only be successfully integrated by spawning Python subprocesses for inference, which adds operational and maintenance complexity, as well as potential performance impacts, particularly for latency-sensitive applications.
All three formats should therefore be avoided in non-Python applications.
Instead, we recommend using ONNX in such cases, as the official JavaScript runtime made the integration a straightforward process.
This underscores ONNX's flexibility and portability across languages and frameworks compared to other formats.
As an alternative, the TensorFlow export format can be used together with the TensorFlow.js runtime module, e.g., if there is a hard constraint on using the TensorFlow framework for model training.
However, this will require the discussed changes to the training and export process using the TFJS Python module, which adds a bit of configuration complexity.
    
A partial explanation of these results may be the different serialization approach of the formats.
ONNX and TensorFlow serialize the models as graphs, which makes them more portable across frameworks and languages.
Joblib and Pickle, on the other hand, serialize models as binaries, which led to substantial integration challenges without an ML framework like \texttt{sklearn} that is aligned with this practice.
Regarding PyTorch, it was surprising that no official JavaScript inference support existed for TorchScript models.
However, PyTorch's recommended way for portability seems to be to either use ONNX\footnote{\url{https://pytorch.org/docs/stable/onnx.html}} or to rely on their ExecuTorch engine\footnote{\url{https://pytorch.org/executorch-overview}} for edge inference, e.g., on Android phones.
However, even the latter does not provide support for JavaScript-based applications.

\textbf{Complex Projects with Increased Preprocessing Needs}: For projects with extensive preprocessing requirements, the TensorFlow export format proved especially valuable, as it can encapsulate both the model and preprocessing logic in a single export file, eliminating the need for additional system modifications.
The ONNX, Pickle, and Joblib formats required exporting preprocessing logic as a separate file, necessitating modifications to runtime modules, which can further complicate the setup.
PyTorch, while capable of handling preprocessing, initially encountered obstacles, which we resolved by training a parallel model with \texttt{sklearn}.
For projects with complex preprocessing, we therefore recommend considering TensorFlow's integrated approach to managing model and preprocessing logic in a single file.
However, one potential downside of this to consider is that model and preprocessing logic can no longer evolve independently and are always replaced together.

Table~\ref{tab:usage-guidance} summarizes our usage recommendations per model export format.
While these can help guide format selection, they are primarily intended to make using the format easier if switching to a different export format is no option.

\section{Threats to Validity}
Several potential threats to validity need to be mentioned.

\subsection{Internal Validity}
One threat for qualitative case studies related to internal validity is the possibility of subjective researcher bias.
While field notes can capture rich and detailed experiences, their creation and interpretation is inherently tied to the background and perception of the individual researchers.
Having a well-designed structured field note template can partially mitigate this, but never fully resolve it.
As a second measure, we discussed and scrutinized the case study results and final ratings within the research team, which led to refinements for inconsistent or unclear parts based on our consensus.
However, it is possible that the results would be slightly different if other researchers had implemented these cases, with potentially other or more encountered challenges or a slight change in the perceived difficulty.
Overall, we do not believe that the general derived guidelines would differ fundamentally, though.

\subsection{External Validity}
The generalizability of case study results is often limited due to the small number of analyzed cases, sometimes just a single unit of analysis.
The comparative nature of our embedded case study, with 30 units of analysis, certainly covered more ground in this regard, and was also grounded in industry-relevant formats and technologies through our preliminary survey.
However, while the chosen examples reflect common practices, ML development methods can vary in industry, which may influence outcomes.
Our two chosen systems were relatively straightforward non-industry applications, which  had more of a synthetic than a real-world character.
Additionally, our findings may be less applicable to very complex systems based on extremely large models, e.g., generative ones.
Expanding sample diversity in future studies should address this limitation.
Another issue could be that we did not study a wide variety of languages and frameworks due to the associated effort.
While we implemented each system in three versions, the similarity between JavaScript and TypeScript may have reduced the generalizability more than including, e.g., Java, C++, or Go.
Including more distinct frameworks or languages in future research could further validate the findings and offer insights into different integration challenges.

\section{Conclusion}
Selecting the right ML model export format is crucial for smooth integration into ML-enabled systems.
Our case study suggests that the ONNX and TensorFlow formats often perform well across diverse environments, with ONNX especially noted for its portability.
Each format demonstrates specific strengths and limitations, with integration quality often depending on the complexity of system needs and configurations.
In particular, we observed that complex systems may encounter integration challenges, notably with preprocessing and system-specific configurations, emphasizing the nuanced requirements for effective deployment.
All formats benefit from strong documentation and community support, which aids integration. 
To promote transparency and reproducibility, we make our study artifacts available online~\footnote{\url{https://doi.org/10.6084/m9.figshare.27613212}}.
Future research should study the integration and maintenance effort of these formats across additional technology stacks and with much larger models, but also their impact on system-level quality attributes such as energy consumption, reliability, performance efficiency, or security.
Additionally, including non-Python ML frameworks like Spark MLlib\footnote{\url{https://spark.apache.org/mllib}} or Caffe\footnote{\url{https://caffe.berkeleyvision.org}} in the study would shed light on the integration efforts in this space.


\footnotesize
\bibliographystyle{IEEEtranN}
\bibliography{source}

\end{document}